\documentclass[prb,preprint]{revtex4-1}
\usepackage{amsmath}  % needed for \tfrac, \bmatrix, etc.
\usepackage{amsfonts} % needed for bold Greek, Fraktur, and blackboard bold
\usepackage{graphicx} % needed for figures

\begin{document}

\title{Transfer Matrix of Scatterers Connected in Parallel}

\author{Yu Jiang}
\email{jiang@xanum.uam.mx}
\affiliation{Departamento de Fisica, Universidad Autonoma Metropolitana - Iztapalapa, Mexico}

\date{\today}

\begin{abstract}
Transport phenomena in parallel coupled scatterers are studied by transfer matrix formulism. We derive a simple recurrence relation for transfer matrix of one-dimensional two-terminal systems consisting of $N$ arbitrary scattering unit cells connected in parallel. For identical scattering sub-units we find that the effects of parallel connection on transport properties of the coupled system can be described by a similarity transformation on the single scatterer, with the similar matrix determined by the scattering matrix of the junction. While for distinct single scatterers, the similar matrices depend on both scattering properties of individual elements and structure of connection topologies.
\end{abstract}


\maketitle % title page is now complete

Transport property of coupled one-dimensional systems is of interest in understanding many physical processes such as charge transport in disordered superconductors \cite{Alex}, Anderson transition in disordered wires \cite{Anderson, Mello}, quantum Hall systems \cite{Klesse}, super-lattices \cite{Vidal}, quantum wires \cite{Exner}, mesoscopic quantum systems\cite{Kowal, Texier},and optical devices\cite{Ali}. The man features of one-dimensional transport can be extracted from the transfer and scattering matrix formulation, in the context of waveguide theory\cite{Xia, Griff, Shao} and the tight-binding formalism\cite{Kowal, CH Wu, Wu}. The $Green^,s$ function approach is another powerful method, particularly in dealing with transport on a network\cite{Andradea}. In one dimension the study of scattering properties of $N$ cells connected in series is considerably simplified with use of transfer matrices\cite{Hua, Sprung, Pereyra, Griff}. In the case of identical cells coupled in series, it has been shown that the "N"-cell transfer matrix can be expressed in terms of single-cell transmission and reflection amplitude via Chebychev polynomials\cite{Sprung, Griff}. This simple formalism allows a straightforward analysis of the whole systems based on the knowledge of the Bloch phase as well as the single cell transport properties.

However, one-dimensional transport phenomena of parallel connected systems have not received much attention, except one-dimensional normal metal rings, or two-lead Aharonov-Bohm rings.  Those simple parallel coupled systems show quite a large variety of interesting quantum interference features\cite{Shapiro, Buttiker, Gefen, Datta, Cahay}. Other parallel coupled systems, like quantum conduction on a $N$-ary Cayley trees, have also been discussed, where one incoming wave is split into $N$ outgoing branches. Note that there are several works on quantum transport properties through some scaled "black boxes" scatterers, where several scatterers are combined into a single re-normalized complex scattering unit, such as the glued binary Carley trees\cite{Shapiro, Liu, Yu}, with use of scattering matrix method in the form of recurrence relation.  But a general theory on the transport through parallel circuits in a tunnel junction is still lacking.

The purpose of this work is to develop a general transfer matrix formalism for parallel coupled two-terminal scattering cells. We derive an exact closed-form expression for the identical cells, which can also be regarded as a recurrence relation for scaled scattering systems, like glued n-ary Cayley trees. By properly choosing the parameters in our formalism we can recover the existing results on one-dimensional quantum rings.

To start with, let us consider a composed scatterer formed by joining two leads of $N$ two-terminal scatterers to the splitting (input) and the merging (output) junctions, respectively, as depicted in Fig.1.
Denote the wave function at the splitting junction in terms of its right- (u) and left-going (v) wave,
\begin{equation}
\psi(x) = u(x) + v(x)
\end{equation}
where  $u(x)=Ae^{ikx}$ and $v(x)=Be^{-ikx}$. Similarly, one has, at the merging point,
\begin{equation}
\psi'(x) = u'(x) + v'(x)
\end{equation}
where  $u'(x)=A'e^{ikx}$ and $v'(x)=B'e^{-ikx}$. Then, the transfer matrix in this representation, which relates the wave state at $O$ to that at $O'$ is given by \cite{Sprung}
\begin{equation}
\begin{bmatrix}
	u \\
	v \\
\end{bmatrix}
=M
\begin{bmatrix}
	u' \\
	v' \\
\end{bmatrix},
\end{equation}

At the junction $O$, we assume that the incoming wave state is scattered uniformly to the $N$ channels, then the wave functions at both sides of the splitter are related by the $(N+1) \times (N+1)$ scattering matrix, defined as
\begin{equation}
\begin{bmatrix}
	v \\
	u_1 \\
	u_2\\
	.\\
	. \\
	. \\
	u_N \\
\end{bmatrix}
%=\bar S
=
\begin{pmatrix}
	 \alpha & \beta & \beta & . & . &. & \beta \\
     \beta & s_{1,1} & s_{1,2} & . & . &. & s_{1,N}\\
     \beta & s_{2,1} & s_{2,2} & . & . &. & s_{2,N}\\
     .\\
     .\\
     .\\
      \beta & s_{N,1} & s_{1,2} & . & . &. & s_{N,N} \\
\end{pmatrix}
\begin{bmatrix}
	u \\
	v_1 \\
	v_2\\
	.\\
	. \\
	. \\
	v_N \\
\end{bmatrix},
\end{equation}
From Eq.(4) it follows that
\begin{eqnarray}
v & = & \alpha u+\beta \sum_{j=1}^{N}v_j , \\
u_i & = & \beta u+\sum_{j=1}^{N}s_{i,j}v_j,  \quad (i=1,2,...,N) .
\end{eqnarray}

On the other hand at the merging junction $O'$ we obtain a couple of similar equations
\begin{eqnarray}
u' & = & \alpha' v'+\beta' \sum_{j=1}^{N}u'_j , \\
v'_i & = & \beta' v'+\sum_{j=1}^{N}s'_{i,j}u'_j,  \quad (i=1,2,...,N) .
\end{eqnarray}
We further assume that on the i-th connecting path between $O$ and $O'$, the transfer matrices are given by
\begin{equation}
M_i=
\begin{pmatrix}
	 m^i_{1,1} & m^i_{1,2} \\
	m^i_{2,1} & m^i_{2,2} \\
\end{pmatrix},
\end{equation}
may represent the scattering process on the $i-$th pathway, which means that $M_i$ contains all transport information of that branch. In view of $Det(M_i)=1$, we have
\begin{equation}
M^{-1}_i=
\begin{pmatrix}
	 m^i_{2,2} & -m^i_{1,2} \\
	-m^i_{2,1} & m^i_{1,1} \\
\end{pmatrix},
\end{equation}
which represents the following state transfer relation
\begin{equation}
\begin{bmatrix}
	u_i \\
	v_i \\
\end{bmatrix}
=M_i
\begin{bmatrix}
	u'_i \\
	v'_i \\
\end{bmatrix},
\end{equation}

The main idea here is to find out an equivalent channel that results from taking into account of the scattering features of all transfer lines connecting the two joining points and reduces the parallel coupled problem to an equivalent serially coupled configuration, that describes the transport from the left lead, through the central "dressed" channel, and to the right lead. To this end we resort to the same technique used by Buttker, by summing and subtracting those relationes obtained from the scattering matrices of the junctions. Therefore, by summing over $i$ in Eq.(6), and together with Eq.(5), we get the following matrix equation:
\begin{equation}
\begin{bmatrix}
	u\\
	v \\
\end{bmatrix}
=\sum_{j=1}^{N}U_j
\begin{bmatrix}
	u_j \\
	v_j \\
\end{bmatrix},
\end{equation}
here we have introduced
\begin{equation}
\gamma_j=\sum_{i=1}^{N}s_{i,j}
\end{equation}
and
\begin{equation}
U_j=\frac{1}{N\beta}
\begin{pmatrix}
	 1 & -\gamma_j \\
	 \alpha & N\beta^2-\alpha\gamma_j \\
\end{pmatrix}
\end{equation}

Following the same procedure, one finds,
\begin{equation}
\begin{bmatrix}
	u'\\
	v' \\
\end{bmatrix}
=\sum_{j=1}^{N}U'_j
\begin{bmatrix}
	u'_j \\
	v'_j \\
\end{bmatrix},
\end{equation}
where
\begin{equation}
\gamma'_j=\sum_{i=1}^{N}s'_{i,j}
\end{equation}
and
\begin{equation}
U'_j=\frac{1}{N\beta}
\begin{pmatrix}
	 N\beta'^2-\alpha'\gamma'_j & \alpha' \\
	 -\gamma'_j & 1 \\
\end{pmatrix}
\end{equation}

Eq.(12) shows how the wave states $(u,v)$ to the left of the splitter is related to the central, transit variables $(u_i,v_i)$ near the diverging junction $O$. On the other hand, Eq.(15) gives the relation between wave states on both sides of the converging point $O'$. Now we derive an equation that connect $(u_i,v_i)$ to $(u_j,v_j)$.
From Eqs(6) and (8) it follows
\begin{eqnarray}
u_i-u_j & = & \sum_{k=1}^{N}s_{i,k}v_k-\sum_{l=1}^{N}s_{j,l}v_l, \quad  (i,j=1,2,3,...,N), \\
v'_i-v'_j & = & \sum_{k=1}^{N}s'_{i,k}u'_k-\sum_{l=1}^{N}s'_{j,l}u'_l,  \quad (i,j=1,2,...,N)
\end{eqnarray}
Inserting (9) into (17) we find

\begin{equation}
Q_i
\begin{bmatrix}
	u_i\\
	v_i \\
\end{bmatrix}
- Q_j
\begin{bmatrix}
	u_j\\
	v_j \\
\end{bmatrix}
=
\sum_{k\neq i, k\neq j}^{N}
\Gamma_{ij,k}
\begin{bmatrix}
	u_k \\
	v_k \\
\end{bmatrix}
\end{equation}
where
\begin{equation}
Q_i=
\begin{pmatrix}
	 1 & -s_{i,i}+ s_{j,i}\\
	 -m^i_{2,1}-(s'_{i,i}-s'_{j,i})m^i_{2,2} & m^i_{1,1}+(s'_{i,i}-s'_{j,i})m^i_{1,2} \\
\end{pmatrix}
\end{equation}
and
\begin{equation}
\Gamma_{ij,k}=
\begin{pmatrix}
	 0 & s_{i,k}- s_{j,k}\\
	 (s'_{i,k}-s'_{j,k})m^k_{2,2} & (s'_{i,k}-s'_{j,k})m^k_{1,2} \\
\end{pmatrix}
\end{equation}
Let us choose $K$-th branch as our reference channel, and introduce
\begin{equation}
a_{i,j}=Q_j-\Gamma_{iK,j}, \quad
\psi_j=
\begin{bmatrix}
	u_j \\
	v_j \\
\end{bmatrix}
\end{equation}
thus, we can rewrite (18) in the following form
\begin{equation}
\sum_{j \neq K}^{N}a_{i,j}\psi_j=Q_i\psi_K
\end{equation}
Solutions to the above linear equations may be expressed as
\begin{equation}
\begin{bmatrix}
	u_i \\
	v_i \\
\end{bmatrix}
=L_{i,K}
\begin{bmatrix}
	u_K \\
	v_K \\
\end{bmatrix}
\end{equation}
where $L_{i,K}$ are $2\times 2$ matrices defined by
\begin{equation}
L_{i,K}=
\sum_{j \neq K}^{N}a^{-1}_{i,j}Q_j
\end{equation}
Note that $A=\{a_{i,j}\}$, and its inverse $A^{-1}=\{a^{-1}_{i,j}\}$ are $2(N-1)\times 2(N-1)$ matrices. The inverse matrix  $A^{-1}$ may be not straightforward to evaluate, in general, nevertheless, it can be shown that for a large class of physical scattering matrices, the matrix $A$ may exhibit simple structure and its inverse can be obtained analytically.
By introducing
\begin{equation}
T_K=\sum_{j=1}^{N}U_jL_{j,K}
\end{equation}
and
\begin{equation}
T'_K=\sum_{j=1}^{N}U'_jM_j^{-1}L_{j,K}M_K
\end{equation}
We finally obtain the transfer matrix for array of scatterers in parallel connection,
\begin{equation}
M =T_KM_KT'^{-1}_K
\end{equation}
This is the main result of this work. It reveals an interesting feature, characteristic of parallel connection, that the transmission property of the whole system is related to that of its member cell, by a unitary transformation. The formulation developed here allows a simple, direct evaluation of the transmission and reflection amplitude, in many symmetric transport processes, as will be demonstrated in later examples. It is worthwhile to pointing out that Eq.(38) may be regarded as a recurrence relation and used for hierarchical composite scatterers, such as Cayley trees with N-ary branches.

Now suppose that our splitter and converter are described by an symmetric $(N+1)$-terminal junction. The scattering matrix is given by
\begin{equation}
s_{i,i} = s'_{i,i}=a, \quad s_{i,j}=s'_{i,j} = b
\end{equation}
where $a$ and $b$, together with $\alpha$ and $\beta$, which are determined by the Lie algebra $SU(N+1)$\cite{Shapiro},
\begin{eqnarray}
\alpha^2+N\beta^2 & = & 1; \\
\alpha+a+(N-1)b & = & 0; \\
\beta^2+a^2+(N-1)b^2 & = & 1; \\
\beta^2+2ab+(N-2)b^2 & = & 0,
\end{eqnarray}
and
\begin{equation}
\gamma_j = \sum_{i=1}^{N}s_{i,j}=a+(N-1)b=-\alpha.
\end{equation}
With such a special scattering matrix for both junctions considered in this work, we have
\begin{equation}
\Gamma_{ij,k}=0
\end{equation}
which leads to the coupling matrix that relates the arbitrary $i$-th and $j$-th channel, as follows
\begin{equation}
L_{i,j}=Q^{-1}_iQ_j
\end{equation}

Now we focus our attention to more specific examples.

(A): Identical scatterers $M_i = M_j$.
From (37) it follows immediately that
\begin{equation}
L_{i,k}=I
\end{equation}
and therefore,
\begin{equation}
T_K=T'_K=\frac{1}{\beta}
\begin{pmatrix}
	 1 & \alpha \\
	 \alpha & 1  \\
\end{pmatrix}
\end{equation}

For free propagation, the transfer matrix on internal branch is given by
\begin{equation}
M_j=
\begin{pmatrix}
      e^{-ikL} & 0 \\
	  0 & e^{ikL}  \\
\end{pmatrix},
\end{equation}

we obtain the transfer matrix
\begin{equation}
M= \frac{1}{1-\alpha^2}
\begin{pmatrix}
	  e^{-ikL}-\alpha^2 e^{ikL} & \alpha( e^{ikL}- e^{-ikL}) \\
	  -\alpha( e^{ikL}- e^{-ikL}) & e^{ikL}-\alpha^2 e^{-ikL}  \\
\end{pmatrix}
\end{equation}

From Eq.(41) we observe that the perfect transmission occurs when $k=n\pi/L$, with $n=0, \pm 1,\pm 2, ...$, which is independent of the coupling configuration. This result is in sharp contrast with the transmission resonance observed for $N$ scatterers coupled in series, wher the perfect transmission appears whenever $|t|^2=1$, and there are additional $N-1$ possibilities when $N$ such scatterers are coupled in series. But in the case of parallel connection, the original transmission resonance does not appear in coupled system.

(B): Quantum ring with two leads.

For $N=2$ and assuming that the scattering matrices are the same for splitter and converter,  Eqs. (27) and (28) read
\begin{equation}
T_2=U_1L_{1,2}+U_2L_{2,2},\quad
T'_2=U'_1L'_{1,2}+U'_2L'_{2,2}
\end{equation}
where $U_1=U_2$, $U'_1=U'_2$ and $L_{2,2}=L'_{2,2}=I$. Now we take
\begin{equation}
M_1=
\begin{pmatrix}
      e^{-ikL_1} & 0 \\
	  0 & e^{ikL_1}  \\
\end{pmatrix},
\quad
M_2=
\begin{pmatrix}
      e^{-ikL_2} & 0 \\
	  0 & e^{ikL_2}  \\
\end{pmatrix},
\end{equation}
with $L=L_1+L_2$ being the ring round length, and it follows
\begin{equation}
Q_1=
\begin{pmatrix}
     1 & 1 \\
	 e^{ikL_1} & e^{-ikL_1}  \\
\end{pmatrix},
\quad
Q_2=
\begin{pmatrix}
      1 & 1 \\
	 e^{ikL_2} & e^{-ikL_2}
\end{pmatrix},
\end{equation}
we obtain
\begin{equation}
L_{1,2}=Q^{-1}_1Q_2=\frac{1}{e^{-ikL_1} - e^{ikL_1}}
\begin{pmatrix}
	 e^{-ikL_1} - e^{ikL_2} & e^{-ikL_1} - e^{-ikL_2} \\
	 -e^{ikL_1} + e^{ikL_2} & -e^{ikL_1} + e^{-ikL_2} \\
\end{pmatrix}
\end{equation}

Assume now that the scattering matrix is given by $\alpha=a=-1/3$ and $\beta=b=2/3$, and introduce
\begin{equation}
\lambda_1= e^{-ikL_1}-e^{ikL_1}, \quad \lambda_2= e^{-ikL_2}-e^{ikL_2}, \quad \lambda_{12}=e^{-ikL_1}-e^{ikL_2}
\end{equation}
we find then
\begin{equation}
L_{1,2}+I=\frac{1}{\lambda_1}
\begin{pmatrix}
	 \lambda_1+\lambda_{12} & \lambda_1+\lambda^*_{12} \\
	 \lambda_1-\lambda_{12} & \lambda_1-\lambda^*_{12} \\
\end{pmatrix}
\end{equation}
Here $\lambda^*$ denotes the complex conjugate of $\lambda$. By a similar procedure we arrive at
\begin{equation}
L'_{1,2}+I=\frac{1}{\lambda_1}
\begin{pmatrix}
	 \lambda_1-\lambda^*_{12} & \lambda_1-\lambda_{12} \\
	 \lambda_1+\lambda^*_{12} & \lambda_1+\lambda_{12} \\
\end{pmatrix}
\end{equation}
Finally we find the transfer matrix for a quantum ring given by
\begin{equation}
M=T_2M_2T'^{-1}_2.
\end{equation}
or
\begin{equation}
M=\frac{1}{1-\alpha^2}
\begin{pmatrix}
    1 & \alpha \\
	\alpha & 1 \\
\end{pmatrix}
(L_{1,2}+I)M_2(L'_{1,2}+I)^{-1}
\begin{pmatrix}
	 1 & -\alpha \\
	 -\alpha & 1 \\
\end{pmatrix}
\end{equation}
After some tedious manipulations, we obtain
\begin{equation}
M_{1,1}=\frac{9e^{-ikL_1}e^{-ikL_2}+e^{ikL_1}e^{ikL_2}-e^{-ikL_1}e^{ikL_2}-e^{ikL_1}e^{-ikL_2}-8}{4(e^{-ikL_1}-e^{ikL_1}+e^{-ikL_2}-e^{ikL_2})}
\end{equation}
\begin{equation}
M_{1,2}=\frac{3e^{-ikL_1}e^{-ikL_2}+3e^{ikL_1}e^{ikL_2}+e^{-ikL_1}e^{ikL_2}+e^{ikL_1}e^{-ikL_2}-8}{4(e^{-ikL_1}-e^{ikL_1}+e^{-ikL_2}-e^{ikL_2})}
\end{equation}
Here we recover the transmission and reflection amplitudes reported in Ref.[\cite{Xia}]. As for the Aharonov-Bohm ring the transport properties can achieved with use of the following the single transfer matrices are given by
\begin{equation}
M_1=
\begin{pmatrix}
      e^{ik_1L_1} & 0 \\
	  0 & e^{-ik_2L_1}  \\
\end{pmatrix},
\quad
M_2=
\begin{pmatrix}
      e^{ik_2L_2} & 0 \\
	  0 & e^{-ik_1L_2}  \\
\end{pmatrix}.
\end{equation}

In summary, an transfer matrix approach for transport on parallel connected scatterers has been derived. It is demonstrated that for identical scatterers connected in parallel the transport properties are determined by that of single cell through a simple recurrence relation. In sharp contrast with serially coupled scattering units, the perfect transmission will be destroyed by parallel coupling, and moreover there are no additional transmission resonances due to the connection.

\begin{figure}[h!]
\caption{Schematic picture of the parallel connected system. The arrows denote the wave propagating directions on each channel at vicinity of the junctions.}
\end{figure}

\end{document}